  \RequirePackage{fix-cm}
  \documentclass[twocolumn]{svjour3}          % twocolumn
  \smartqed  % flush right qed marks, e.g. at end of proof
  \usepackage{amssymb}
  \usepackage{tabularx}
  \usepackage[utf8]{inputenc}
  \usepackage{graphicx}
  \usepackage{hyperref} %ハイパーリンクを作成する事ができる
  \hypersetup{% hyperrefオプションリスト
  setpagesize=false,
  bookmarksnumbered=true,%
  bookmarksopen=true,%
  colorlinks=true,%
  linkcolor=blue,
  citecolor=red,
  }
  \usepackage{textcomp} %テキスト上に角度記号などを入力するためのパッケージ
  \usepackage{mathptmx}      % use Times fonts if available on your TeX system

  \journalname{ArXiv}
  
\begin{document}
  
  \title{Deep learning-based topological optimization for representing a user-specified design area%\thanks{Grants or other notes
  }
  
  
  \author{Keigo Nakamura         \and
          Yoshiro Suzuki* %etc.
  }
  
  
  \institute{F. Author: Keigo Nakamura \at
  Tokyo Institute of Technology, Department of Mechanical
  Engineering, 2-12-1 Ookayama, Meguro-ku, Tokyo 152-8552, Japan. \\
                Tel.: +813-5734-3178\\
                Fax: +813-5734-3178\\
                \email{nakamura.k.br@m.titech.ac.jp}           %  \\
             \and
             *: Corresponding author: Yoshiro Suzuki \at 
             Tokyo Institute of Technology, Department of Mechanical
             Engineering, 2-12-1 Ookayama, Meguro-ku, Tokyo 152-8552, Japan. \\
             \email{ysuzuki@ginza.mes.titech.ac.jp} 
  }
  
  \date{}

  \maketitle
  
  \begin{abstract}
  Presently, topology optimization requires multiple iterations to create an optimized structure for given conditions.
  Among the conditions for topology optimization, the design area is one of the most important for structural design.
  In this study, we propose a new deep learning model to generate an optimized structure for a given design domain and
  other boundary conditions without iteration. For this purpose, we used open-source topology optimization MATLAB code
  to generate a pair of optimized structures under various design conditions. The resolution of the optimized structure
  is 32 \texttimes \  32 pixels, and the design conditions are design area, volume fraction, distribution of external forces, and
  load value. Our deep learning model is primarily composed of a convolutional neural network (CNN)-based encoder and
  decoder, trained with datasets generated with MATLAB code. In the encoder, we use batch normalization (BN) to
  increase the stability of the CNN model. In the decoder, we use SPADE (spatially adaptive denormalization)
  to reinforce the design area information. Comparing the performance of our proposed model with a CNN model
  that does not use BN and SPADE, values for mean absolute error (MAE), mean compliance error,
  and volume error with the optimized topology structure generated in MATLAB code were smaller,
  and the proposed model was able to represent the design area more precisely.
  The proposed method generates near-optimal structures reflecting the design
  area in less computational time, compared with the open-source topology optimization MATLAB code.
  \end{abstract}
  
  \section{Introduction}
  \label{sec:1}
  In recent years, significant improvements in deep learning algorithms and 
  computer hardware have made it possible to apply deep learning to medical 
  imaging and speech recognition tasks. The most remarkable improvement in 
  deep learning algorithms is the convolutional neural network (CNN), which 
  is particularly well suited to image recognition. CNNs have achieved high 
  performance in tasks such as making and synthesizing images, recognizing 
  and classifying subjects, completion of perforated images, removing noise, 
  and generating high resolution images from low resolution images.
  
  Several studies show that deep learning could be applied not only to 
  imaging and language applications, but also in mechanical fields, such 
  as fluid simulation (Kim et al. \hyperlink{Kim}{2019}), and structure optimization 
  (Yu et al. \hyperlink{Yu}{2019}). Such research demonstrates the ability of CNNs to reduce 
  the cost of computational time.
 
  Structural optimization is a method for optimally designing structures 
  to maximize the target performance under imposed conditions, or design areas.
 
  Structural optimization is categorized into three types: Size optimization 
  (Pragert \hyperlink{Pragert}{1974}; Svanberg \hyperlink{Svanberg}{1982}), 
  shape optimization (Ding \hyperlink{Ding}{1986}; Haslinger 
  and Mäkinen \hyperlink{Haslinger}{2003}), and topology optimization (Bendsøe and Kikuchi \hyperlink{Bendsoe1988}{1988}; 
  Bendsøe \hyperlink{Bendsoe1989}{1989}). In size optimization, structure is optimized by changing 
  only the length and the thickness, without altering the shape of the structure. 
  In shape optimization, the outer shape of the structure is changed in addition 
  to length and thickness. Compared with size optimization, shape optimization 
  provides greater freedom, and can produce optimized structures with higher 
  performance than is achievable with size optimization. In topology optimization, 
  structures are designed by considering the mass density distribution of the 
  material as a design objective. Hence, the size, outer shape, and topology 
  of the structure can be designed. Compared with size and shape optimization, 
  it provides greater freedom in structural design, and can also design optimized 
  structures with higher performance. However, topology optimization requires a 
  large number of design parameters and many updates, which is associated with 
  high computational cost.
 
  When the design area is 2D, the subject of the topology optimization 
  is the material density of each cell arranged in a grid, and displayed 
  as a matrix. In topology optimization, optimizing the material density of 
  all cells determines the shape of the structure, to achieve high strength 
  and rigidity.
 
  On the other hand, an image can be displayed as matrix with luminance 
  values arranged in grid pattern pixels. CNNs are good at extracting the 
  shape of a subject from images, and making the images(matrix information). Therefore, it is 
  possible that CNNs are also good at estimating a structure's rigidity 
  from the material density's matrix, and also making the material density's matrix. 
  There have been several studies aiming to reduce the computational cost of topology 
  optimization using CNNs. 
 
  CNN models were applied to topology optimization by Sosnovik et al. (\hyperlink{Sosnovik}{2017}), 
  who reported success in reducing computational cost. The SIMP method 
  (Bendsøe and Sigmund \hyperlink{Bendsoe1999}{1999}), which is a conventional form of topology 
  optimization, requires many updates to the material density distribution 
  during structural optimization. While the optimization process is ongoing, 
  the structure at that point (i.e. the structure before reaching its final 
  optimized form) is defined as the intermediate structure. The difference 
  between the pre- and post-update structure at a given step in the optimization 
  process is defined as the gradient. The CNN model proposed by Sosnovik et al. 
  takes the intermediate structure and gradient as inputs, and outputs the final 
  optimized structure. However, since the inputs to Sosnovik's CNN model do not 
  include information on the design area or boundary conditions, these parameters 
  cannot be directly reflected in the structure output from the model.
 
  In \hyperlink{Banga}{2018}, Banga et al. used a CNN to reduce the computational cost of 3D 
  topology optimization. However, as with Sosnovik's approach, their model 
  uses the intermediate structure and gradient as inputs to the CNN.
 
  Since these two methods depend on the solution obtained from the SIMP method, 
  they do not achieve topology optimization using a CNN alone. Rather, both 
  methods use a CNN to reduce the number of material density updates required 
  to complete optimization.
 
  In \hyperlink{Rawat}{2018}, Rawat and Shen proposed a method for obtaining topology-optimized 
  structures (final material density distribution) without iteration. Their 
  method is based on a generative adversarial network (GAN) (Goodfellow et al. \hyperlink{Goodfellow}{2014}) 
  model. The GAN model is composed of two deep learning (DL) models, and is 
  trained by competition between the two models; a form of adversarial learning. 
  In the GAN model proposed by Rawat and Shen, the DL models are both CNNs, hence 
  the GAN model outputs a topology-optimized structure using CNNs only, without 
  any input from the SIMP method. However, constraints such as volume fraction 
  and boundary conditions, which are important for performing topology optimization, 
  cannot be specified in this CNN model.
 
  Conversely, with the DL model proposed by Yu et al. (\hyperlink{Yu}{2019}), it is possible 
  to specify a point where external force is applied, the external force's 
  direction, the structure's volume constraint, and the position of the fixed 
  point of the structure. Yu's DL model is composed of a CNN model and a GAN model, 
  and can output topology-optimized structures based on the above conditions. 
  Not only does Yu's CNN model achieve topology optimization without iteration, 
  it allows specification of conditions for topology optimization.
 
  As described above, some research exits into CNN-based topology optimization 
  without iteration. However, to the best of our knowledge, no previous CNN 
  models exist that support specification of the design area for topology 
  optimization. All the studies mentioned above use a predetermined design 
  area that cannot be changed. In other words, these CNN models can only output 
  an optimized structure with a design area of the same shape and size. 
  Design area is an important condition for performing topology optimization. 
  Having a higher degree of freedom for the design area increases the versatility 
  of the optimization method.
 
  In this paper, we propose a CNN model that takes the distribution of external 
  forces, value of external forces, volume fraction, and design area as inputs, 
  and outputs the material density distribution without iteration. The output 
  has the same rigidity as a structure optimized using conventional methods.
 
  To output the optimized structure, our CNN model uses SPADE ResBlk 
  (spatially adaptive denormalization residual block) (Park et al. \hyperlink{Park}{2019}); 
  a technique for generating images by applying a semantic segmentation mask 
  to a CNN. Park et al. created a GAN model using SPADE. A semantic 
  segmentation mask depicting silhouettes of the sky, sea, mountains, clouds, 
  soil, trees, etc. in different colors is input to this GAN model via SPADE 
  ResBlk, and an image of the actual scene based on the mask is output by the 
  GAN model. SPADE ResBlk is described in detail in Section \ref{sec:2.2}.
 
  Design area and external force distribution, which are important conditions 
  for topology optimization, can be regarded as image information. For example, 
  if the design area is divided into a grid, and the area where materials are 
  arranged is represented by black (1), while the area where materials are not 
  arranged is represented by white (0), the design area becomes an image composed 
  of black and white. In the case of external distribution, if a luminance value 
  is set in a place where an external force is applied, and the luminance values 
  depends on the external load values, the external force distribution becomes a 
  grayscale image. It is expected that using SPADE ResBlk, the mask (design area 
  and external force distribution) expressed in image format can be passed to the 
  CNN accurately. Using SPADE ResBlk makes it possible for the design area and 
  external force distribution to be accurately represented in the optimized 
  structure (optimal material density distribution) output by the CNN. 
 
  The most novel aspect of this research is the development of a CNN model 
  that can support specification of design area information for topology 
  optimization. Having compared and verified our model against the CNN model 
  mentioned above (Yu et al. \hyperlink{Yu}{2019}), our CNN model was found to output a material 
  density distribution that more accurately reflects the volume constraint, 
  with greater rigidity. In addition, our model more accurately represents 
  the design area, and outputs a material density distribution closer to the 
  optimal structure obtained using conventional topology optimization (SIMP method). 
  The results of this comparison are described in detail in Section \ref{sec:4}.

  \section{Related research}
  \label{sec:2}
  \subsection{The CNN model proposed by Yu et al. (2019)}
\begin{figure*}
      \includegraphics[width=1\textwidth]{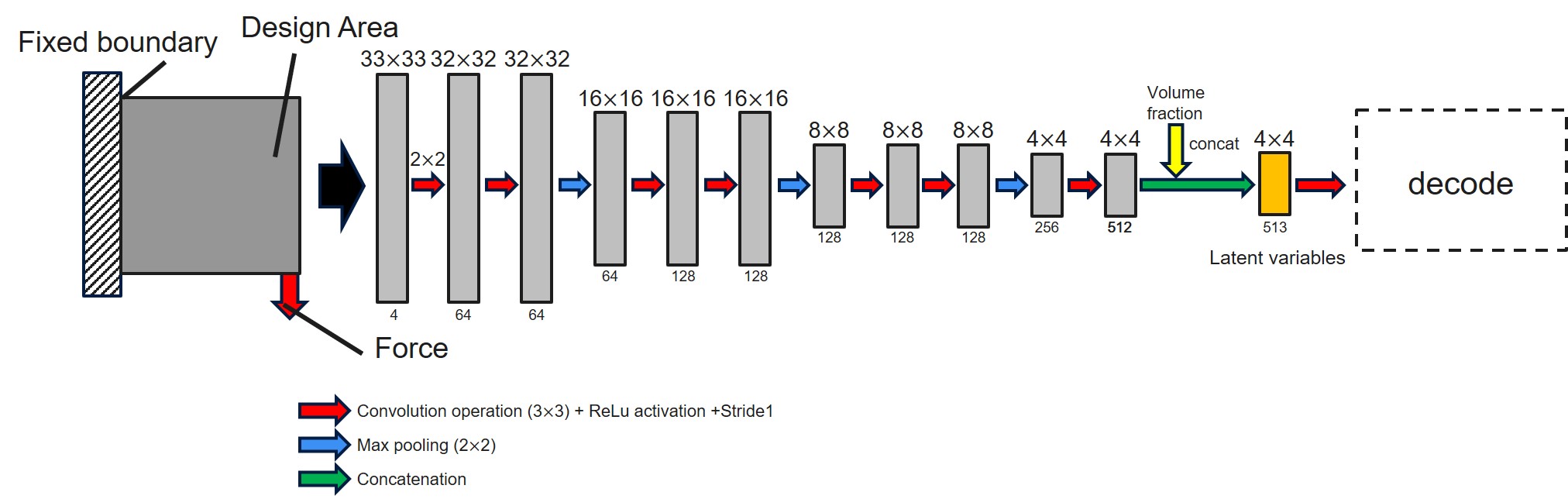}
    \caption{Architecture of Yu's encoder CNN model (Yu et al. 2019)}
    \label{fig:Yu_encoder_CNN} 
    \end{figure*}

\begin{figure*}
        \includegraphics[width=1\textwidth]{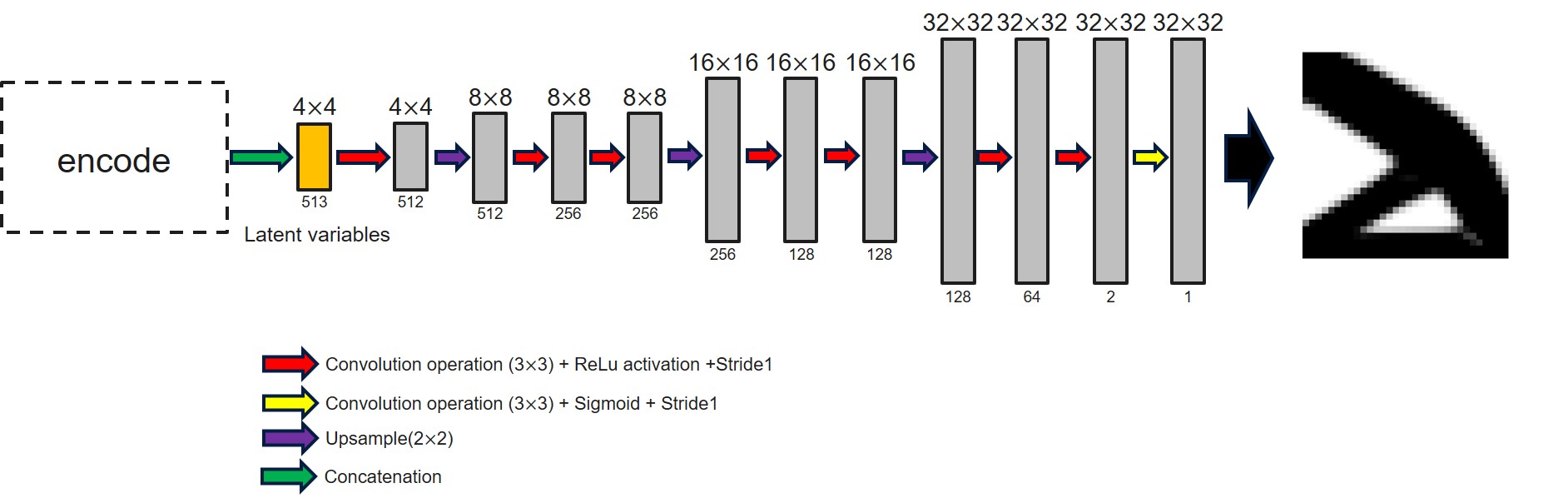}
      \caption{Architecture of Yu's decoder CNN model (Yu et al. 2019)}
      \label{fig:Yu_decoder_CNN} 
    \end{figure*}

  The DL model for topology optimization proposed by Yu et al. (\hyperlink{Yu}{2019}) 
  is composed of a CNN model and a GAN model.The CNN model is able to accept 
  specification of a point where external force is applied, the external 
  force's direction, the volume constraint of the structure, and the 
  position of the fixed point of the structure; its output is a 
  topology-optimized material density distribution of 32 \texttimes \  32 pixels.
  The GAN model upscales the resolution of the material density distribution 
  output by the CNN model from 128 \texttimes \  128 pixels into 
  32 \texttimes \  32 pixels. This paper focuses on the CNN model that outputs 
  a material density distribution of 32 \texttimes \  32 pixels. Their model consists 
  of a CNN (encoder CNN) (Fig. \ref{fig:Yu_encoder_CNN}) that extracts and 
  compresses the features of the input information, and a CNN (decoder CNN) 
  (Fig. \ref{fig:Yu_decoder_CNN}) that expands the extracted features to 
  32 \texttimes \  32 pixels. 
 
  The input information is: The single point where external force is applied, 
  the external force's direction, the volume constraint of the structure, 
  and the position of the fixed point of the structure.
  The design area is fixed as a square of 32 \texttimes \  32 pixels.
  The conditions are as follows.
    \begin{itemize}
      \item Volume fraction: 0.2–0.8
      \item Position of the fixed point
      \item Application point of single external force: Node 1 to 1089
      \item Direction of external force: 0\textdegree \ to 360\textdegree
      \end{itemize}

  Based on these input conditions, Yu's CNN model outputs a structure 
  (material density distribution) that maximizes rigidity. When generating 
  datasets for training and evaluation using FEM (finite element method), they use MATLAB 
  open-source topology optimization code (Andreassen et al. \hyperlink{Andreassen}{2011}). 
  Yu et al. set a fixed displacement position and external force load position 
  for the nodes, so both the fixed position information and the load external 
  force information are a 33 \texttimes \  33 matrix (whereas the material density 
  distribution is a 32 \texttimes \  32 matrix). 
 
  First, the position of the fixed point and load external force information, 
  in the 33 \texttimes \  33 matrix, are input to the CNN model. 
 
  In the encoder CNN, the input condition features are extracted using a 
  convolutional layer and a max pooling layer, and a feature map of 
  4 \texttimes \  4 pixels is generated. This 4 \texttimes \  4 pixel 
  feature map and volume fraction form the input to the decoder CNN. This decoder 
  CNN gradually enlarges (up-sampling, un-pooling) the input 4 \texttimes \  4 
  pixel feature map, finally outputting a 32 \texttimes \  32 material density 
  distribution. 

  Both the encoder and decoder CNNs use a rectified linear unit (ReLU) 
  function (Nair and Hinton \hyperlink{Nair}{2010}) as their activation function.
  However, in the output layer of the decoder CNN, a sigmoid activation function 
  (Mitchell \hyperlink{Mitchell}{1997}) is used to output a real value of 0 to 1, 
  representing the material density distribution of each element.

  The MAE (mean absolute error) of the difference in material density distribution 
  between the optimized structure obtained using a conventional topology 
  optimization method and that obtained with Yu's method is used as a loss function, 
  and the CNN model is trained to minimize this MAE.
  In addition, an ADAM optimizer (Kingma and Ba \hyperlink{Kingma}{2014}) is used 
  as the learning algorithm.

  \subsection{SPADE (Park et al. 2019)}
  \label{sec:2.2}
\begin{figure}
      \includegraphics[width=84mm]{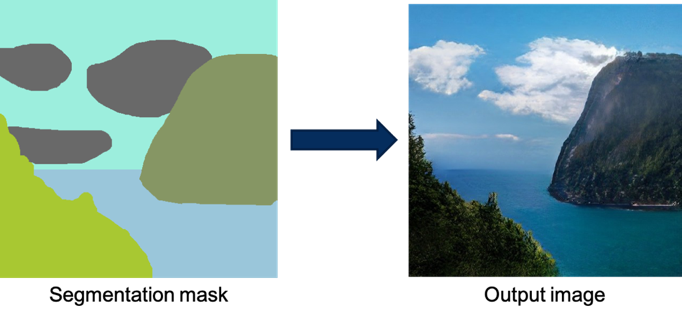}
    % rightとleftを(a)(b)にする予定
    \caption{A realistic image (right) generated from a semantic segmentation mask (left) by GauGAN (Park et al. 2019)(generated by using NVIDIA's demo site (http://nvidia-research-mingyuliu.com/gaugan) by the author of this paper)}
    \label{fig:SPADE_image}
    \end{figure}

  The CNN image generation literature includes several models that use a 
  semantic segmentation mask (mask) as the input to a CNN, 
  outputting an image based on the mask; for example, SIMS 
  (Qi et al. \hyperlink{Qi}{2018}) and pix2pixHD (Wang et al. \hyperlink{Wang}{2018}). 
  In \hyperlink{Park}{2019}, Park et al. proposed a normalization method 
  called SPADE as a method to represent the mask in the output image.
  The GAN model (GauGAN), created by Park et al., uses SPADE. 
  By inputting a mask drawn by coloring the labels of sky, sea, clouds, mountains, forests, etc.,
  GauGAN can generate images that closely approximate real-life scenery, 
  based on the mask, as shown in Figure \ref{fig:SPADE_image}.

\begin{figure}
      \includegraphics[width=84mm]{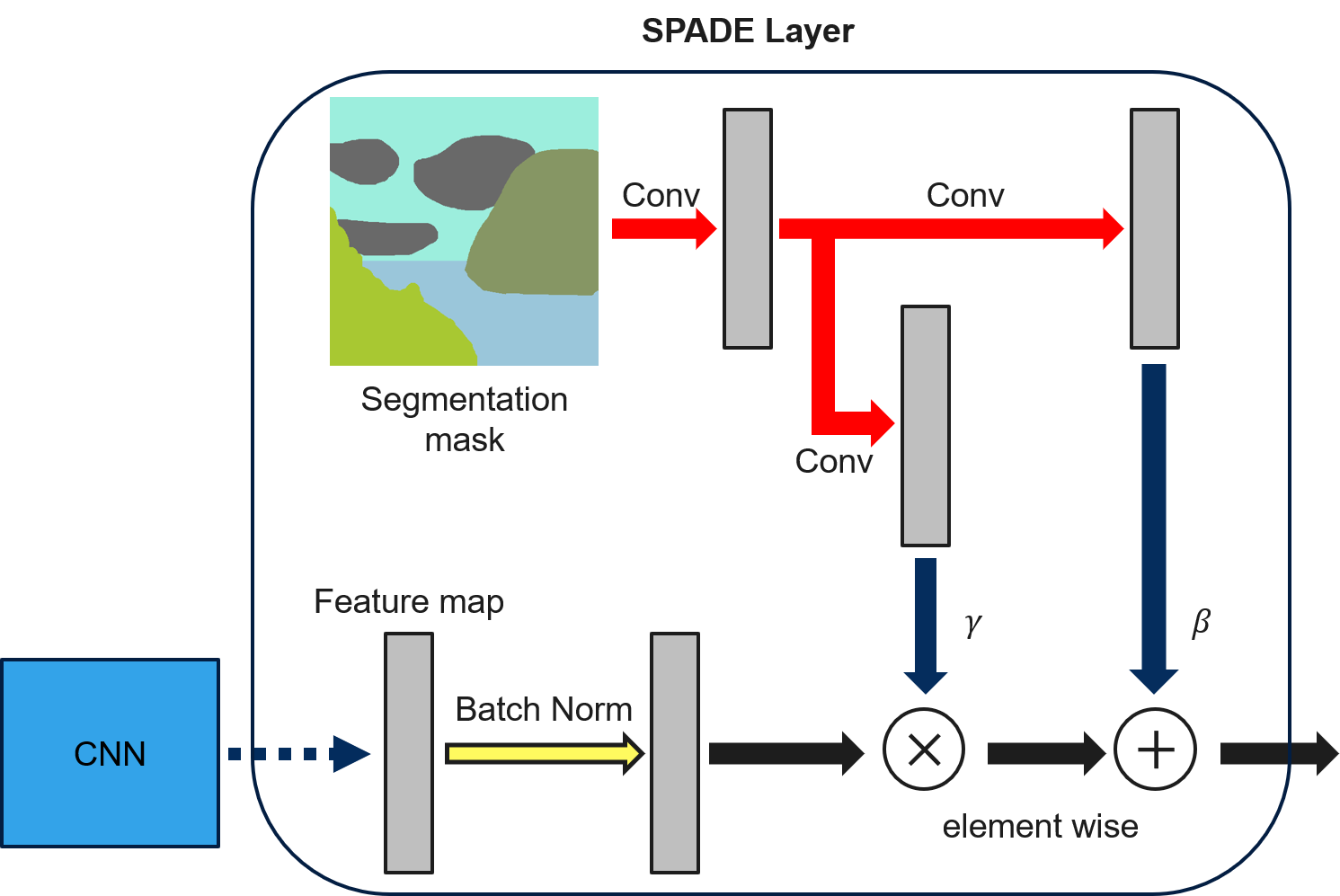}
    \caption{Architecture of SPADE layer(Park et al. 2019)}
    \label{fig:SPADE_layer} 
  \end{figure}

  Figure \ref{fig:SPADE_layer} shows the flow in the SPADE layer.
  Here, the mask is first projected onto the convolution layer to extract a feature map.
  Then, the scale and bias of each pixel of the mask's feature map are obtained through the convolution layer.
  Mask, scale, and bias information is injected into the normalized feature map, which comes from inside the CNN model, 
  and is normalized by performing batch normalization (BN) (Loffe and Szegedy \hyperlink{Loffe}{2015}).
  This is the normalization method performed inside the SPADE layer.
  
  \begin{figure*}
    \centering
    \includegraphics[width=1\textwidth]{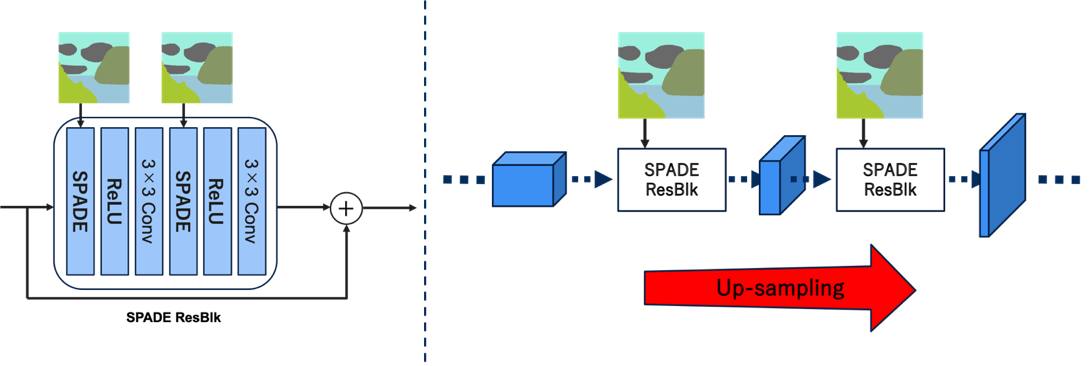}
  \caption{Architecture of SPADE ResBlk(left) and the section of the generator architecture (i.e., the decoder) with the SPADE ResBlks (Park et al. 2019)}
  \label{fig:SPADE_ResBlk} 
  \end{figure*}
  We shall now describe the process in the SPADE layer with calculation formulas.
  Let {\it H, W }  be the number of pixels in the vertical and horizontal directions of the mask.
  Then, let the matrix indicating the mask be 
  \begin{math}
  {\bf m} \in  \mathbb L^{H\times W}
  \end{math}.
  \begin{math} \mathbb L\end{math} is a set of integers, and each integer has label information such as mountains or sky.
  Assuming that the feature map in the \begin{math} i\end{math}-layer of the CNN that
  generates the image is input to the SPADE layer. Let \begin{math} {\bf h}^i\end{math} be the feature map in 
  the \begin{math} i\end{math}-layer with batch size {\it N}, and let \begin{math} {C}^i\end{math}, \begin{math} {H}^i\end{math}, 
  \begin{math} {W}^i\end{math} be the number of channels and pixels in the vertical 
  and horizontal directions that \begin{math} {\bf h}^i\end{math} has. The value of each pixel in the feature map 
  output by the SPADE layer is then given by equation (\ref{eq:1}).
    \begin{equation}
      \label{eq:1}
      {\scriptstyle
      \gamma_{c,y,x}^i({\bf m})\frac{h_{n,c,y,x}^i-\mu_c^i}{\sigma_c^i}+\beta_{c,y,x}^i({\bf m})
       (n \in  N, c \in  C^i,y \in  H^i , x \in  W^i)}
    \end{equation}
  where \begin{math} h_{n,c,y,x}^i\end{math}  are the values of the feature map input to the SPADE layer
  from the \begin{math} i\end{math}-layer, and \begin{math} h_{n,c,y,x}^i\end{math} represents the values before BN.
  \begin{math}\mu_c^i\end{math} and \begin{math}\sigma_c^i\end{math} are the mean and standard deviation values, 
  used to normalize \begin{math} h_{n,c,y,x}^i\end{math} for each channel in the BN process. 
  \begin{math}\mu_c^i\end{math} and \begin{math}\sigma_c^i\end{math} are derived from equations (\ref{eq:2}) and (\ref{eq:3}).
    \begin{equation}
      \label{eq:2}
      \mu_{c}^i=\frac{1}{NH^iW^i}\sum_{n,y,x} h_{n,c,y,x}^i
    \end{equation}
    \begin{equation}
      \label{eq:3}
      \sigma_{c}^i=\sqrt{\frac{1}{NH^iW^i}\sum_{n,y,x} ({h_{n,c,y,x}^i})^2-{\mu_{c}^i}^2}
    \end{equation}
  The values \begin{math}\gamma_{c,y,x}^i({\bf m})\end{math}  and \begin{math}\beta_{c,y,x}^i({\bf m})\end{math} are 
  learned parameters obtained by inputting masks to two convolution layers in the SPADE layer. 
  The values \begin{math}\gamma_{c,y,x}^i({\bf m})\end{math} and \begin{math}\beta_{c,y,x}^i({\bf m})\end{math} represent scale and bias for each pixel. 
  The values \begin{math}\gamma_{c,y,x}^i({\bf m})\end{math} and \begin{math}\beta_{c,y,x}^i({\bf m})\end{math} are used for scale and bias for the {\it i}-layer's feature map, 
  normalized by BN, and calculated as per equation (\ref{eq:1}). 
  In this manner, the mask information is represented on the {\it i}-layer's feature map at the SPADE layer.

  Park et al. also created the SPADE ResNet Block (SPADE ResBlk) (Figure \ref{fig:SPADE_ResBlk} (left)) by combining the SPADE layer with ResNet (He et al. \hyperlink{He}{2016}). 
  The SPADE ResBlk is used multiple times during the up-sampling process
  (Figure \ref{fig:SPADE_ResBlk} (right)). 
 
  In this study, we developed a CNN model that outputs the optimal material density 
  distribution representing the specified design area, using a SPADE ResBlk that 
  represents the mask information on the feature map. At this point, the masks 
  input to the SPADE ResBlk are design area, volume fraction, and external forces.

  \section{Proposed method}
  \label{sec:3}
  \subsection{User-specified conditions}
  In this study, the conditions for topology optimization solved by the proposed CNN model are as follows.
  \begin{itemize}
    \item Volume fraction
    \item Design area
    \item Distribution and load value of external forces
  \end{itemize}
 
  In the case of Yu's topology optimization CNN model (Yu et al. \hyperlink{Yu}{2019}), 
  the user can specify the volume fraction, position of the external force and direction, 
  and position of the fixed point. However, they cannot specify the shape of the design area. 
  
  In this study, we constructed a CNN model supporting specification of not only the volume fraction, 
  the location of each external force and each load value, but also the design area shape as a topology 
  optimization condition. Our CNN model outputs an optimized topology structure of 32 \texttimes \  32 pixels 
  (the material density distribution with the highest rigidity) under the specified conditions.

  \subsection{CNN model architecture}
  \begin{figure*}
    \includegraphics[width=1\textwidth]{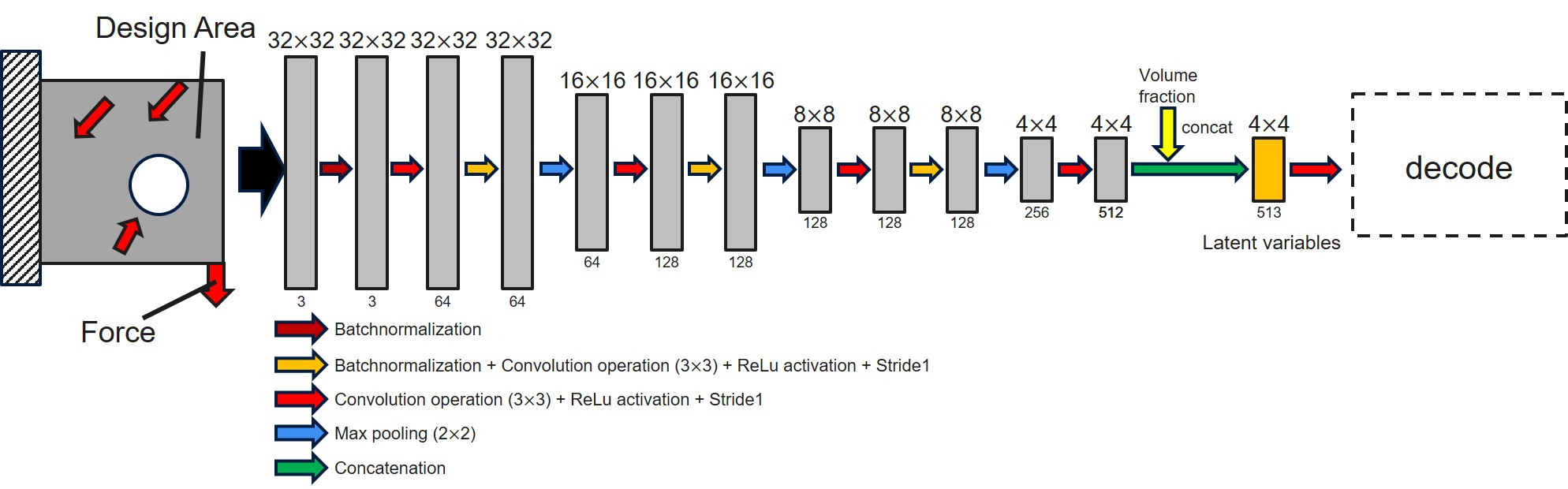}
  \caption{Architecture of our encoder CNN model}
  \label{fig:encoder} 
  \end{figure*}

  \begin{figure*}
    \centering
      \includegraphics[width=0.9\textwidth]{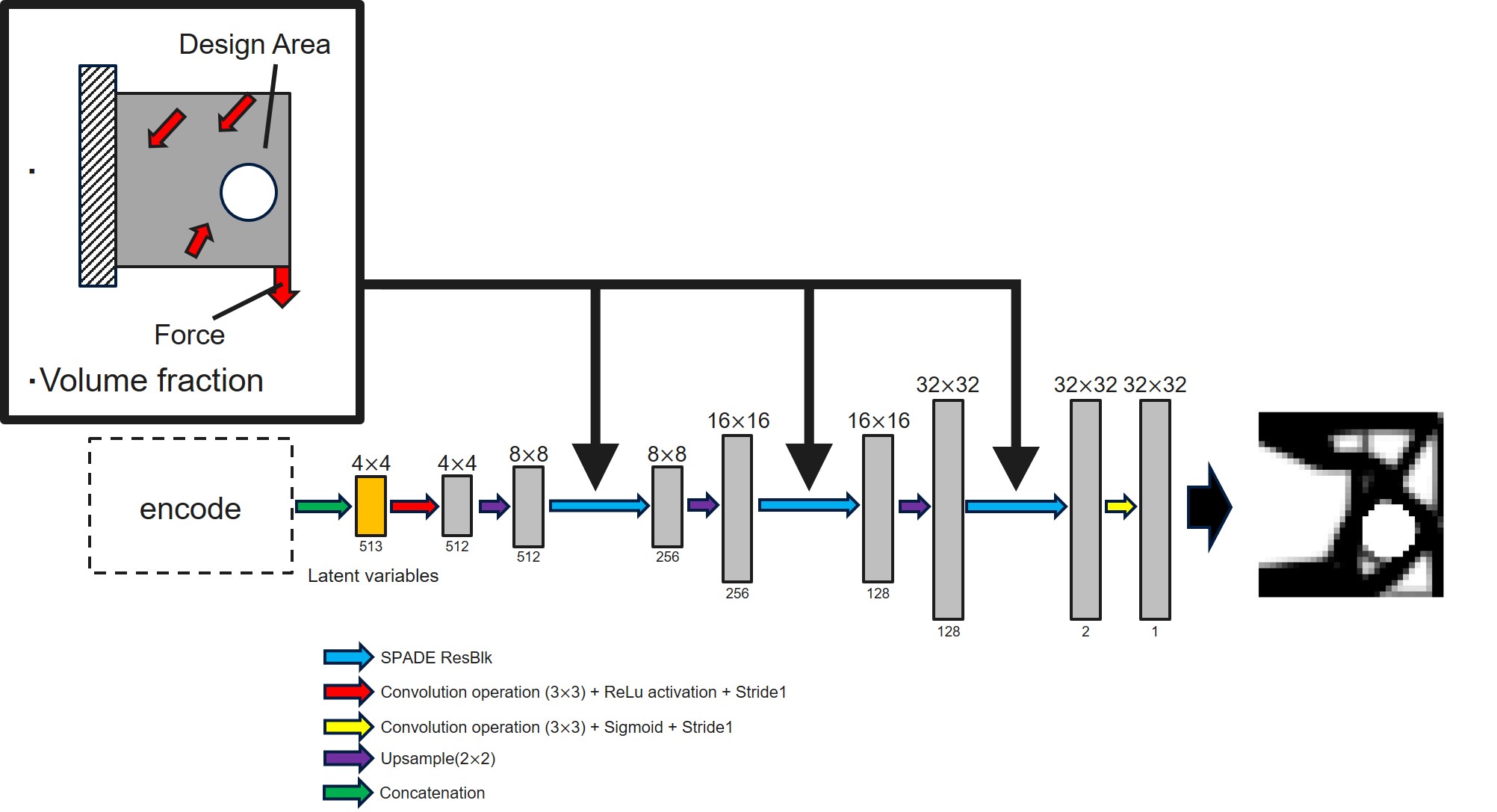}
    \caption{Architecture of our decoder CNN model}
    \label{fig:decoder} 
  \end{figure*}

  \begin{figure}
    \includegraphics[width=84mm]{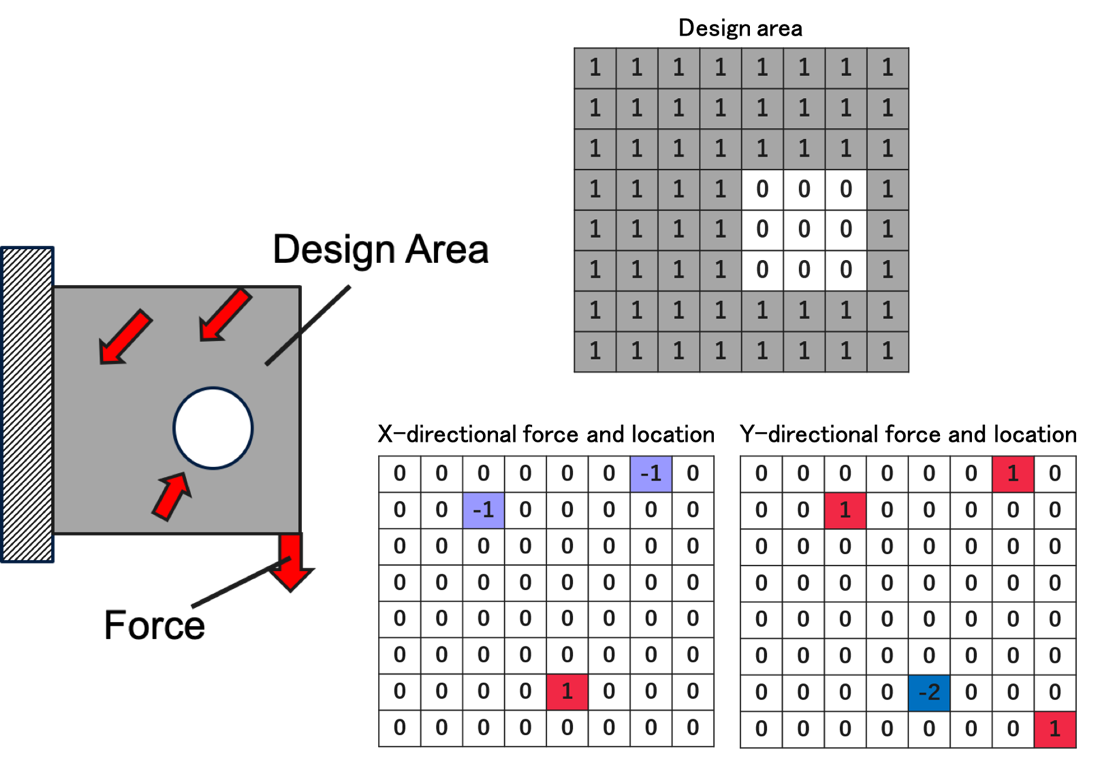}
  \caption{Example of the format of the design area and external force information input into the encoder CNN model}
  \label{fig:input}
  \end{figure}

  \begin{figure}
    \includegraphics[width=84mm]{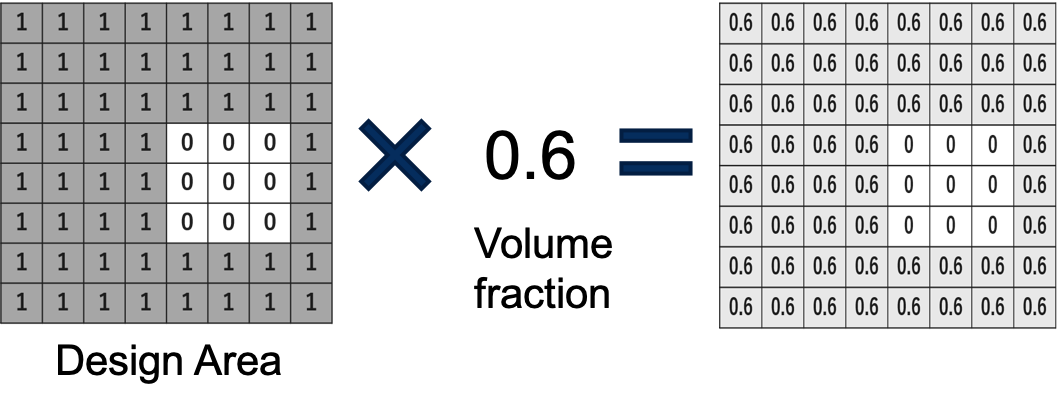}
  \caption{Example of the format combining both the design area and maximum volume fraction information as an input to the SPADE ResBlk (Park et al. 2019)}
  \label{fig:designformat}
  \end{figure}

  Figures \ref{fig:encoder} and \ref{fig:decoder} show the architecture of our CNN model, 
  based on the encoder and decoder CNNs in Yu's model.
  
  The first information input into the encoder CNN is the design area and external force conditions 
  (external force load location and load value) (Fig. \ref{fig:input}). The shape of the design area is represented 
  by a 32 \texttimes \  32 matrix. In other words, the shape of the design area is represented by a matrix where the 
  design area is 1 and the material non-placeable area is 0. The external force condition is represented 
  by two matrices of 32 \texttimes \  32. Each component of the first matrix represents the load value in the {\it x} direction 
  of the external force applied to each element (in FEM, each component represents the element force), and 
  the second matrix represents the load value of the external force in the {\it y} direction.

  In the encoder CNN, at the time of input and during the process of compressing information, 
  the BN normalization process is performed multiple times. The information input into the encoder CNN 
  is compressed and finally output as a 4 \texttimes \  4 feature map. This feature map and volume fraction information 
  are input into the decoder CNN. Specifically, the feature map output by the encoder CNN and a matrix in 
  which all 4 \texttimes \  4 components are volume fraction values are concatenated in the channel direction, and the 
  processed 4 \texttimes \  4 feature map is input to the decoder CNN. 

  The decoder CNN gradually expands the 4 \texttimes \  4 input information (feature map), using SPADE ResBlk (Park et al. \hyperlink{Park}{2019}) 
  to reinforce the area information, before finally generating the optimal 32 \texttimes \  32 material density distribution. 
  The information (feature map) input into the decoder CNN includes optimization conditions (a matrix representing 
  the design area, and a matrix representing the external force distribution, and volume fraction). However, some 
  of the input information may be lost during feature map expansion. To counter this information loss, 
  SPADE ResBlk is used not only to represent the area information, but also to reinforce the optimization conditions. 
  First, a matrix representing the design area, and a matrix representing the external force distribution and volume 
  fraction information are input into SPADE ResBlk; based on the resulting output, the feature maps of multiple 
  layers in the decoder CNN are denormalized. The volume fraction information (scalar information) is combined 
  with the matrix representing the design area information to form the SPADE ResBlk input. The volume fraction 
  and the design area information are combined by multiplying the volume fraction value and the matrix representing 
  the design area where each component is 0 or 1, as shown in Figure \ref{fig:designformat}; the resulting matrix is input into the 
  SPADE ResBlk. In other words, the matrix, including volume fraction and design area information, and two matrices 
  representing the external force distribution are input into the SPADE ResBlk. The output layer of the decoder 
  CNN uses a sigmoid activation function (Mitchell \hyperlink{Mitchell}{1997}) so that the material density (real value of 0 to 1) can 
  be output.

  \section{Evaluation of the effectiveness of the proposed method}
  \label{sec:4}
  \subsection{Conditions of the dataset for training and evaluating our CNN model}
  We set the following topology optimization conditions and used them to evaluate 
  the effectiveness of the proposed method in this study. 
  \begin{itemize}
    \item Volume fraction: 0.2–0.8 (same as Yu's study (Yu et al. 2019))
    \item Design area: The design area is a square, with one circle inside. This circle represents the area where the material cannot be placed. The radius of the circle can be specified as smaller than 25\% of the length of the square. The center of the circle is located within the square design area of 32 \texttimes \  32 pixels. To determine whether our model could output an optimized structure with a complex shape, we selected a curved circle instead of a polygon (e.g. a rectangle) as the shape of the area where the material cannot be placed.
    \item Fixed boundary condition: Fix the left edge of the design area
    \item External forces' distribution and load value: External forces are located in the design area. The maximum number of external load points is 4, and the load values are selected from five patterns, \textminus2, \textminus1,0,1,2 in x and y directions, respectively. The load value is discrete (integer) rather than continuous.
  \end{itemize}

  \subsection{Method to generate the training, evaluation, and test dataset}
  Based on the conditions described in the previous section, we generated training, 
  evaluation, and test datasets using the SIMP method (Bendsøe and Sigmund \hyperlink{Bendsoe1999}{1999}), 
  which, as previously noted, is a conventional topology optimization method. 
  We used MATLAB open-source topology optimization code (Andreassen et al. \hyperlink{Andreassen}{2011}), 
  in which topology optimization is performed using FEM based on the SIMP method. 

\begin{figure}
    \includegraphics[width=84mm]{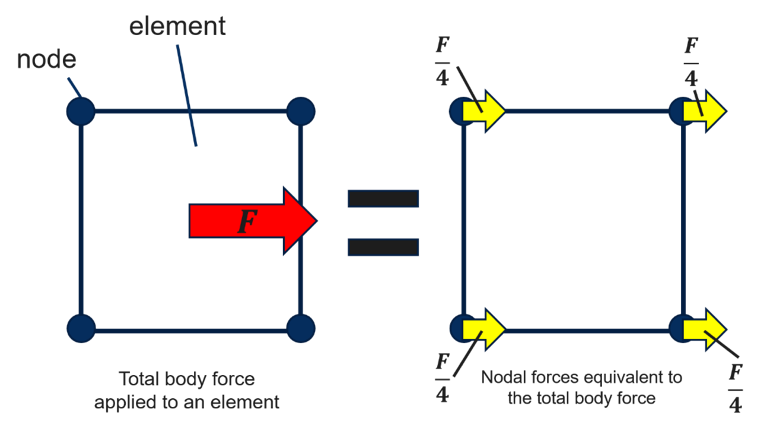}
  \caption{Spliting the body force into four nodal forces with a quarter of the body force's value 
  when converting body force into nodal forces}
  \label{fig:force}
  \end{figure}

  When performing FEM, the structure or design space to be analyzed is divided into elements. 
  The conditions of the force applied to each element are supplied, and the analysis is performed. 
  There are two types of forces in FEM: Element forces and nodal forces. 
  Element force refers to the force applied to the entire element (Figure \ref{fig:force} (left)), 
  whereas nodal force refers to the force applied to a node. Nodes represent the points 
  that comprise an element, and four nodes form a square element. The force applied to 
  this node is called the nodal force (Figure \ref{fig:force} (right)). As shown in figure \ref{fig:force}, 
  when defining the force on one element, a single value is used to represent the element force. 
  Conversely, when defining the nodal force on one element, four values are needed. 
  From the above, the number of values required for setting the external force condition 
  varies depending on the type of force, element force, and nodal force. In this case, 
  the design area is composed of a matrix of 32 \texttimes \  32 pixel elements, and the nodes that comprise 
  this design area are represented by a 33 \texttimes \  33 matrix.

  In this study, the design conditions input into the CNN included external force conditions 
  and design areas. When inputting an external force condition into the CNN as a nodal force, 
  the input matrix will be a 33 \texttimes \  33 matrix, which differs from the 32 \texttimes \  32 matrix showing the 
  design area information, and indicating the element information. Since it is difficult to 
  input matrices of different sizes to the CNN simultaneously, it is inconvenient for the matrices 
  containing the design area information and external force conditions to be different sizes. 
  Therefore, in our study, the external force condition input into the CNN represents the element 
  force defined for each element. In Andreassen's MATLAB code (\hyperlink{Andreassen}{2011}), when specifying the nodal force, 
  which is external force information, we split the body force into four nodal forces with a quarter 
  of the body force's value, as shown in figure \ref{fig:force} (right).

  Under these rules, we varied the volume fraction, the shape of the design area, 
  and the external force conditions to create a total of 370,000 datasets. 
  The entire collection of datasets was split into training, validation, and test datasets with 
  an 8:1:1 ratio, such that 296,000 datasets were used for learning, 37,000 for verification, 
  and 37,000 for test.

  \subsection{Training and evaluating the effectiveness of proposed method}
\begin{figure}
    \includegraphics[width=84mm]{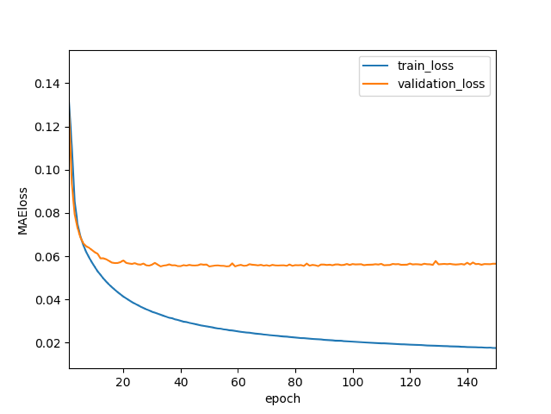}
  \caption{Loss curves for the training and validation datasets of the proposed CNN model}
  \label{fig:losscurve}
  \end{figure}

  Our CNN model was trained under the following learning conditions. 
  \begin{itemize}
    \item Batch size: 512
    \item Loss function: MAE between the material density distribution obtained using the SIMP method our CNN model
    \item Learning algorithm: ADAM
    \item Learning rate: 0.01
  \end{itemize}
  Our CNN model was trained using the above conditions. 
  Figure \ref{fig:losscurve} shows the value of MAE loss for each epoch during training. 
  We evaluated the speed of calculating the material density distribution 
  between our CNN model trained as above, and the SIMP method 
  (Andreassen et al. \hyperlink{Andreassen}{2011}) (Table \ref{tab:1}). The time required to calculate the 
  optimal material density distribution was 0.07634 seconds per case using 
  our model, and 0.4393 seconds per case for the SIMP model, using the same CPU. 
  In other words, our trained CNN model demonstrated an 83\% reduction in 
  calculation time versus the conventional SIMP method.

  \begin{table}
    \centering
    \caption{Comparison of computational time between the conventional 
    topology optimization method and the proposed method}
    \label{tab:1}
    \begin{tabular}{|c||r|r|} \hline 
     & \begin{tabular}{c}
      {\scriptsize SIMP method}\\{\scriptsize (Andreassen et al. 2011)}
      \end{tabular} & {\scriptsize Proposed method} \\ \hline
      \begin{tabular}{c}
        {\scriptsize Average}\\{\scriptsize computational time}
        \end{tabular}& 0.4393 s & 0.07634 s \\ \hline
    \end{tabular}
  \end{table}

  Next, to compare the effects of SPADE ResBlk and BN, 
  we trained the proposed model without BN, and Yu's CNN model, 
  which is equivalent to the proposed model without BN and SPADE ResBlk.

  \subsubsection{Results for validation datasets}
  Validation error results (minimum value of MAE for the validation 
  datasets while training) for each model are as follows.
  \begin{itemize}
    \item Yu's CNN model: 0.112 (115 epochs)
    \item Proposed CNN model (Without BN): 0.059 (37 epochs)
    \item Proposed CNN model: 0.055 (51 epochs)
  \end{itemize}
  The number in parentheses indicates the epoch number when the 
  validation error reached the minimum value. The proposed method 
  with SPADE ResBlk achieved a smaller MAE loss value than Yu's CNN 
  model, thus confirming the effectiveness of SPADE ResBlk. Comparing 
  our proposed CNN model with and without BN, the MAE loss value was smaller 
  with BN. Results for the validation datasets therefore show that the CNN 
  model incorporating both SPADE ResBlk and BN was more effective.

  \subsubsection{Results for test datasets}
  \begin{figure}
    \centering
    \includegraphics[width=84mm]{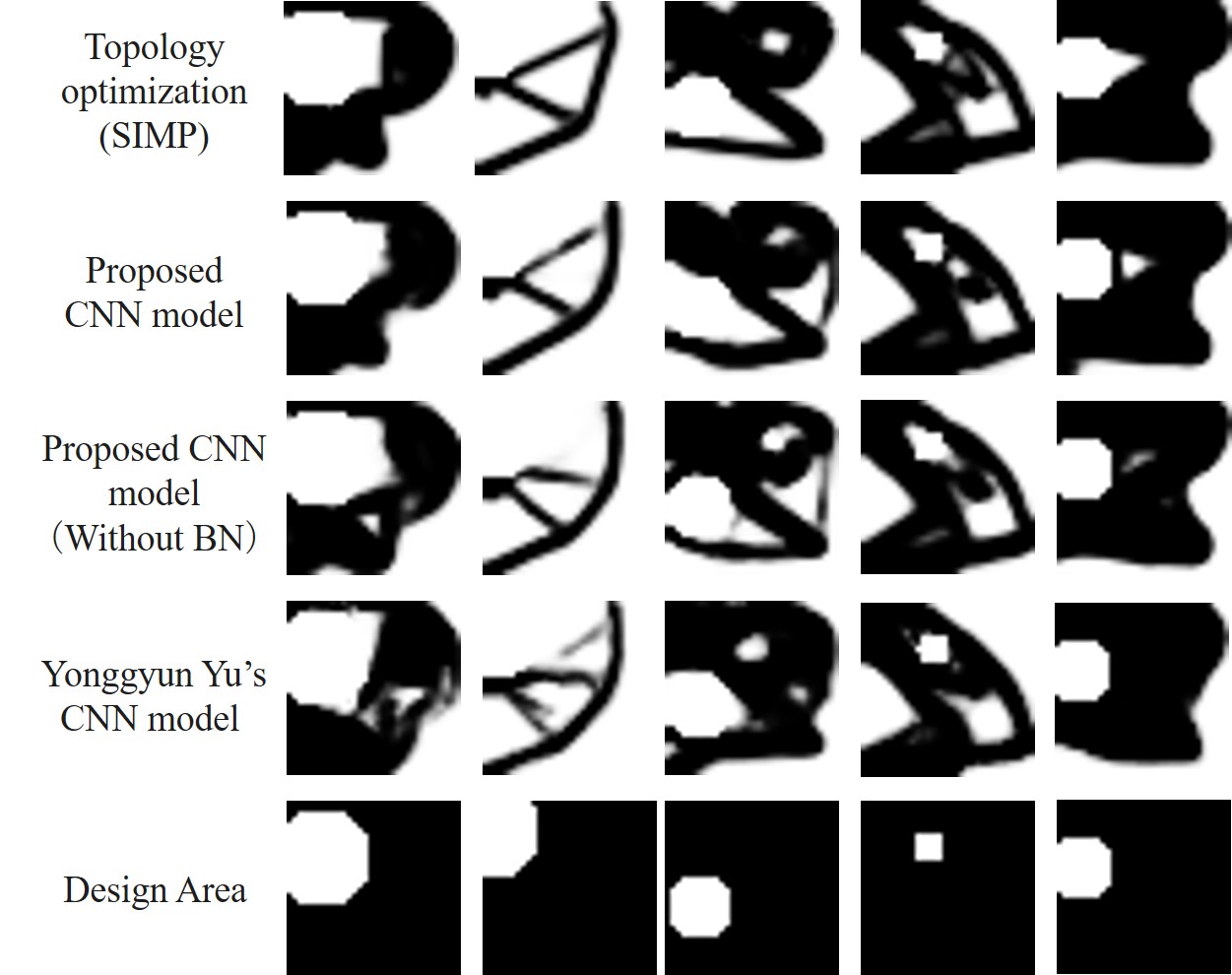}
  \caption{Optimized structures obtained with the various methods 
  and the specified design areas}
  \label{fig:structure} 
  \end{figure}

  \begin{table*}
    \centering
    \caption{Comparison of accuracy between Yu's CNN model 
    and our proposed CNN model (with or without BN)}
    \label{tab:2}
    %\begin{tabularx}{174mm}{|p{28.15mm}||l|l|l|l|} \hline
    \begin{tabular}{|c||r|r|r|r|} \hline
     & \begin{tabular}{c}
      Average error for mean\\compliance (higher 80\%)
      \end{tabular} & 
      \begin{tabular}{c}
        Average error for mean\\compliance (lower 20\%)
        \end{tabular} & 
        \begin{tabular}{l}
          Average error for \\volume fraction
          \end{tabular} & 
          \begin{tabular}{c}
            Rate of the number\\following design area\\(threshold: 0.01)
            \end{tabular}
         \\ \hline 
         Yu's model & 18.7\% &51624956\%&10.28\%&68.2\% \\ \hline
         \begin{tabular}{c}
          Proposed model\\(without BN)
          \end{tabular}
  &3.80\%&424987\%&9.91\%&98.0\%\\ \hline
  Proposed model&2.75\%&103872\%&9.70\%&99.6\%\\ \hline
    \end{tabular}
  \end{table*}

  \begin{figure}
    \includegraphics[width=84mm]{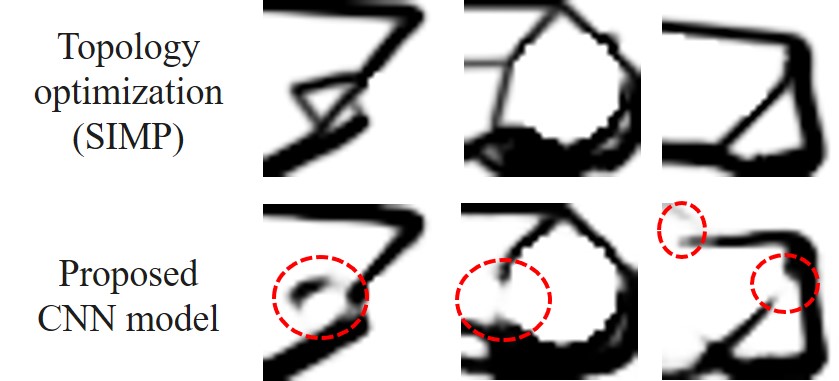}
  \caption{Red circles show examples of structures with disconnection}
  \label{fig:disconnection} 
  \end{figure}
  
  In the section above, we compared the models' performance 
  using the validation datasets. To evaluate performance with 
  the test datasets, we used Yu's CNN model, our proposed CNN model 
  (without BN), and our proposed CNN model when the validation error 
  reached the minimum value. 
  Figure \ref{fig:structure} shows the design area, the respective structures output 
  by each CNN model, and calculated using conventional topology 
  optimization (SIMP method), based on the design area in the test datasets. 
  The following three indices were used for evaluation.
  \begin{itemize}
    \item Average error for mean compliance with optimal 
    material density distribution calculated from the SIMP method
    \item Average error between the specified volume fraction and the 
    volume of the output material density distribution
    \item Ratio of output material density distribution that 
    follows the specified design area
  \end{itemize}
  Table \ref{tab:2} presents the results comparing the three models described above against 
  the three evaluation indices. From the result, the CNN model equipped with 
  SPADE ResBlk and BN had the best performance for all three indices, confirming 
  the validity of our proposed model. 

  In the following section, we shall compare each index in detail. We calculated 
  and compared the mean compliance error between the material density distributions 
  output by each CNN model and calculated using the SIMP method, respectively. 
  Mean compliance is the index relating to rigidity, with low mean compliance 
  indicating high rigidity. Note that each CNN model occasionally outputs a 
  discontinuous material density distribution, as shown in figure \ref{fig:disconnection}. 
  In this case, the discontinuous material density distribution has very low rigidity 
  compared to the optimal material density distribution obtained using the 
  SIMP method, and the mean compliance becomes very large. Therefore, when 
  calculating the average error of the mean compliance, we adopted a method 
  to evaluate what percentage of the average error is in the higher ranked 
  dataset among all the test dataset, assuming that the material density 
  distributions output by CNN which has smaller error is the higher ranked data.

  \begin{figure}
    \includegraphics[width=84mm]{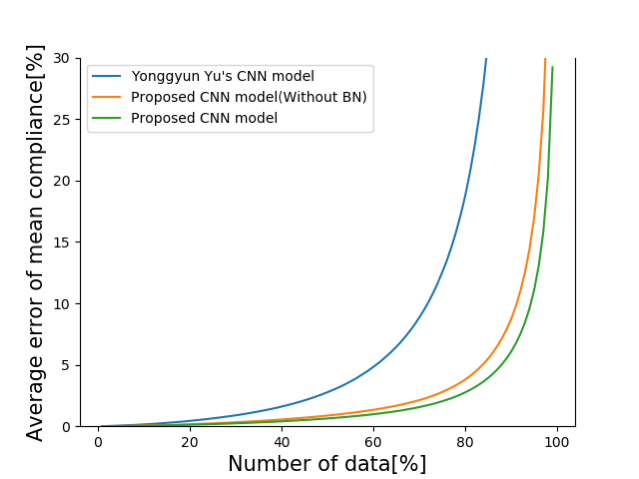}
  \caption{Average error of the mean compliance of the optimized 
  structures for each percentage of higher rank datasets}
  \label{fig:meancompliance}
  \end{figure}

  \begin{figure}
    \includegraphics[width=84mm]{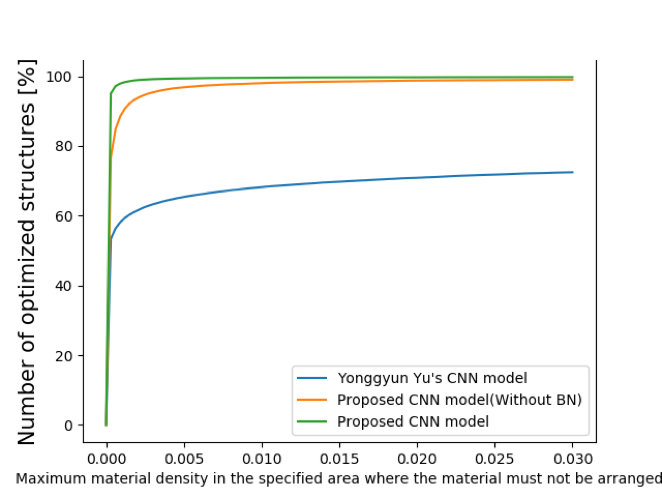}
  \caption{Relationship between the number of optimized structures and the 
  maximum material density in the specified area where the material must 
  not be placed}
  \label{fig:designarea}
  \end{figure}

  Figure \ref{fig:meancompliance} shows the average error for each percentage of higher ranked 
  results for Yu's model, and our proposed model with and without BN, 
  respectively. According to figure \ref{fig:meancompliance}, our proposed model always has the 
  smallest average error, regardless of the ratio of the higher ranked results. 
  Table \ref{tab:2} summarizes the average error for mean compliance in the material 
  density distribution of the top 80\%, as a representative value. 
  For Yu's CNN model, the average error for the top 80\% of results was 18.7 \%, 
  compared to 3.80\% for the proposed model without BN, and 2.75\% for the 
  proposed model with BN, demonstrating that the average error for mean compliance 
  is reduced by using SPADE ResBlk and BN. Table \ref{tab:2} also shows the average error 
  for the lower 20\% of the datasets. Again, the proposed model has the smallest 
  error; about 1/516 of the error of Yu's model. These results indicate that our 
  proposed model can reduce error for the rarely output discontinuous material 
  density distribution. The effectiveness of our proposed model in outputting 
  material density distributions of higher rigidity than Yu's model is thus 
  demonstrated. 

  Next, we compared the models in terms of volume fraction. 
  We used the average volume fraction error between the value provided to the 
  CNN model and the volume of the material density distribution output by the 
  CNN model. The average error of the volume error of the volume fraction in all 
  test datasets was 10.28\% for Yu's CNN model, 9.91\% for the proposed model 
  without BN, and 9.70\% for the proposed model. Our model kept the average error 
  within 10\%. Based on this result, the proposed model clearly outputs a 
  structure that more accurately reflects the volume fraction in topology 
  optimization.

  \begin{figure}
    \includegraphics[width=84mm]{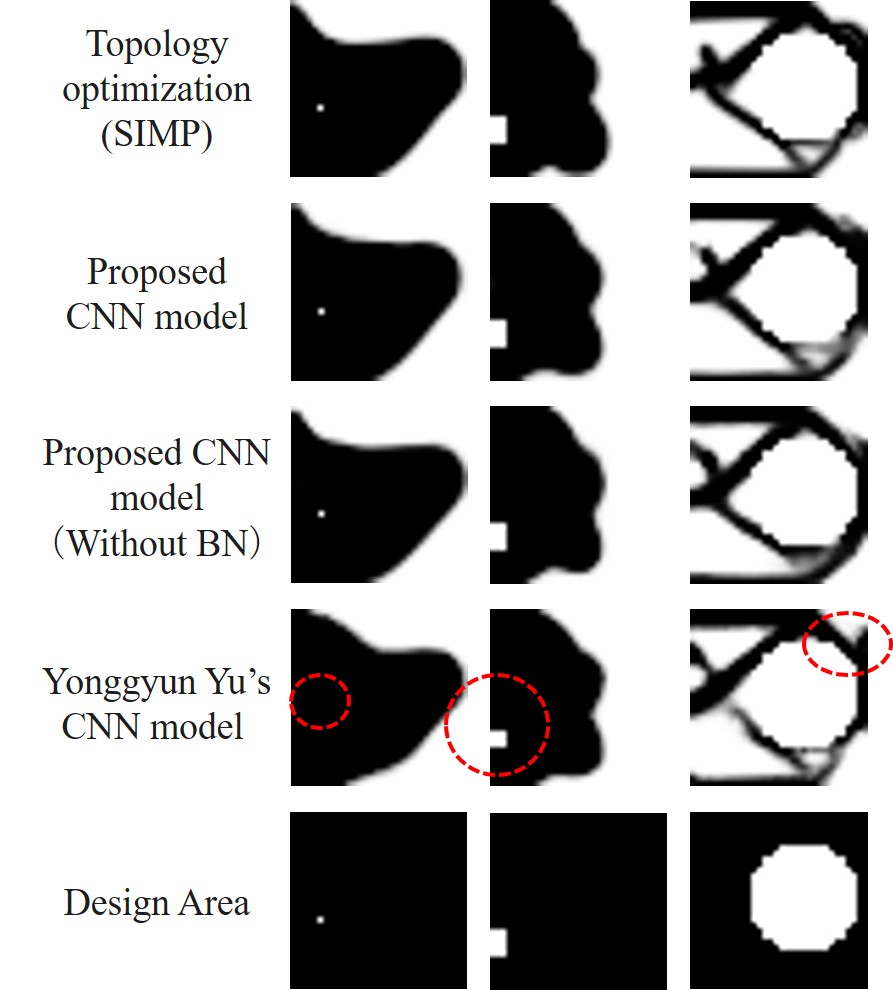}
  \caption{Red circles show the example of Yu's CNN model 
  that did not reflect the specified design areas}
  \label{fig:notreflectdesignarea} 
  \end{figure}

  Finally, we evaluated whether the material density distribution output 
  by each CNN model reflected the shape of specified design area. This evaluation 
  is the most important element of this study. Ideally, the material density 
  should be 0 in the area where the material cannot be placed (circular area 
  in figure \ref{fig:structure} design area). However, since the CNN only has a sigmoid output function, 
  the material density distribution actually output from the CNN model is 
  unlikely to be completely zero. Therefore, we decided that the condition 
  of the design area would be satisfied if the maximum value of the material 
  density in the circular area was below the threshold value. For example, 
  with a threshold value of 0.01, if the maximum value of the material density 
  in the circular area is 0.005, it is lower than the threshold value, and thus 
  the design area condition is satisfied. Figure \ref{fig:designarea} summarizes the percentage 
  of datasets where the design area condition was satisfied for each threshold 
  value in the test datasets. It is apparent that our model achieved a higher 
  ratio of results representing the design area for each threshold value than 
  the other models. In fact, Yu's model output more instances of material density 
  distributions that did not satisfy the design area, as per the example shown 
  in figure \ref{fig:notreflectdesignarea}. Using the data obtained for a threshold value of 0.01 as a 
  representative result, the ratio of the material density distribution 
  representing the design area in each model is 68.2\% for Yu's model, 98.0\% 
  for the proposed model without BN, and 99.6\% for the proposed model. 

  However, it has been confirmed that our proposed model occasionally 
  outputs structures with a discontinuous material density distribution 
  like Yu's CNN model, as shown in figure \ref{fig:disconnection}. In this case, the error of 
  the mean compliance becomes very large. This suggests that using the MAE 
  of the material density distribution as the loss function is not sufficient 
  to train the CNN model. Even if the CNN model is trained to minimize the MAE 
  of the material density distribution, it cannot minimize the error between 
  the mean compliance of the structure output by the CNN model and the mean 
  compliance of the training datasets' structure. To overcome this problem, 
  we propose the following approach, which we plan to investigate in future work.
  \begin{itemize}
    \item Add the mean compliance error to the loss function 
    (setting loss function as the weighted sum of the mean compliance 
    error and the MAE of the material density distribution). 
    \item Make the GAN model incorporating our CNN model output more 
    realistic continuous material density distributions.
  \end{itemize}

  \section{Conclusion}
  \label{sec:5}
  We have proposed a new topology optimization method using a CNN, 
  which is a key technology for DL, currently attracting much attention 
  in the field of image generation. Our CNN model can support specification 
  of important design areas, volume fractions, and external force distribution 
  (load position and load value of external force) as design conditions for 
  topology optimization. No previous CNN model has achieved such topology 
  optimization.

  In a previous study, Yu et al. (Yu et al. \hyperlink{Yu}{2019}) presented a CNN model 
  for topology optimization, capable of outputting the optimized structure 
  (optimal material density distribution) in a single calculation, and also 
  able to accept specified volume fractions, external force conditions, and 
  positions of fixed point. Our model differs from Yu's in its application of 
  BN (Loffe and Szegedy \hyperlink{Loffe}{2015}) and SPADE ResBlk (Park et al. \hyperlink{Park}{2019}).

  Evaluating the material density distribution output by the CNN model for 
  the specified conditions of the shape of the design area, volume fraction, 
  and external force distribution, our model was shown to be highly effective, 
  as follows.
  \begin{itemize}
    \item Compared to Yu's model, our CNN model can output a material 
    density distribution closer to that achieved by the conventional 
    SIMP method (Bendsøe and Sigmund \hyperlink{Bendsoe1999}{1999}). 
    \item Among the material distributions output by our CNN model, at an 80\% 
    material density distribution, the average error of the mean compliance 
    with the material density distribution obtained by the SIMP method is kept 
    to 2.75\%. Therefore, our model can output a material density distribution 
    almost identical to that of SIMP.
    \item 99.6\% of all material density distributions output by our 
    CNN model reflect the shape of the specified design area. In other 
    words, our CNN model does not place material outside the specified 
    design area. 
    \item Computational time needed to obtain the optimal material density 
    distribution using our model was reduced by 83\%, compared with the 
    SIMP method.
  \end{itemize}

    In the future, we would like to improve the accuracy of the material 
    density distributions output by our CNN model by implementing the 
    improvements described at the end of section \ref{sec:4}. Additionally, in 
    this paper, verification was performed for 2D design areas of  
    32 \texttimes \  32 pixels, but we aim to extend our CNN model to support 
    higher resolutions or 3D design areas.

  \end{document}